\documentstyle[11pt,paspconf,epsf]{article}

\begin{document}

\title{The Dynamical Evolution of Galaxies in Clusters}

\author{John Dubinski}
\affil{CITA, University of Toronto, McLennan Labs, 60 St.\ George St.,
Toronto, ON, Canada M5S 3H8; dubinski@cita.utoronto.ca}

% Notice that some of these authors have alternate affiliations, which
% are identified by the \altaffilmark after each name.  The actual alternate
% affiliation information is typeset in footnotes at the bottom of the
% first page, and the text itself is specified in \altaffiltext commands.
% There is a separate \altaffiltext for each alternate affiliation
% indicated above.

%\altaffiltext{1}{Visiting Astronomer, Cerro Tololo Inter-American Observatory. 
%CTIO is operated by AURA, Inc.\ under cooperative agreement with the National
%Science Foundation} 
%\altaffiltext{2}{Society of Fellows, Harvard University} 
%\altaffiltext{3}{Patron, Alonso's Bar and Grill}

% The abstract is entered in a LaTeX "environment", designated with paired
% \begin{abstract} -- \end{abstract} commands.  Other environments are
% identified by the name in the curly braces.

% Poster authors ONLY may omit the abstract in order to gain a little
% more page space for the text of the poster.

\begin{abstract}
The evolution of galaxies is driven strongly by dynamical processes
including internal instabilities, tidal interactions and mergers.
The cluster environment is a useful laboratory for studying these effects.
I present recent results on simulations
of interacting populations of spiral and elliptical galaxies 
in the cosmological collapse of a cluster, showing the formation
of the central, brightest cluster galaxies through merging, the effect of
tidal interactions and merging on galaxy
morphology over cosmic history, and the distribution and kinematics of the
tidal debris field. (See www.cita.utoronto.ca/$\sim$dubinski/rutgers98 for
figures and animations.)
\end{abstract}

% Keywords should be included, but they are not printed in the hardcopy.


\section{Introduction}

In most cosmological scenarios, galaxies form early and rapidly
in the collapse of density perturbations
but continue to evolve over a Hubble time both in luminosity
and morphology.
The ageing of the galaxy stellar population
leads to a gradual dimming of their light, 
a signal that is now picked up
in distant spirals 
and elliptical galaxies (e.g. Vogt et al. 1996; Kelson et al. 1997)
Dynamics primarily drives the galaxy morphological evolution
through: 1) internal gravitational instabilities such as the bar
instability which may influence bulge formation, 2)
tidal interactions which can explain much of the ``disturbed'' and irregular
morphology of galaxies at high $z$ (e.g. Oemler et al. 1997) 
3) the merging of disks to form 
some or most of the ellipticals (Toomre 1977).  Merging obviously
contributes to evolution of the galaxy luminosity function.

Galaxy clusters are the best place to investigate the connection between
dynamics and morphology because they contain a diverse population of galaxies
that have interacted strongly many times over their lifetime.  Dressler (1984)
has aptly described clusters as laboratories of galaxy formation, in
particular in the way they emphasize the importance of gravitational
interactions.
Here, I explicitly follow that lead by setting up numerical experiments 
of galaxy interactions in cosmological clusters.

The simulation of galaxy dynamics in a cosmological context at sufficient
resolution to resolve detailed dynamical effects is now becoming feasible. 
Kiloparsec scales can now be resolved in the volume
surrounding a collapsing cluster. 
Most aspects of the dynamical evolution of
galaxies do not require the complicated details of dissipative 
galaxy formation, so there is no need for expensive hydro calculations.

A simple technique that allows studies of galaxy dynamics in
clusters works as follows.  First, a cosmological N-body 
simulation in a large volume is run
and a cluster size dark halo is identified at $z=0$.
The simulation is re-examined at early times ($z=3$ to 2) and all
dark halos that will end up  in the cluster are replaced with 
N-body models of disk (and possibly elliptical) galaxies scaled according
the mass and circular velocity of the halos with at least
$10\times$ the resolution.  The rationale is that galactic disks 
should form rapidly prior to cluster collapse and the bulk of their
dynamical
evolution will be driven by the interactions they experience when they fall
into the collapsing cluster.  Simulations are then continued with the resolved
galaxy models to $z=0$.

So far I have applied this technique in 2 simulations: the first with a
poor cluster ($\sigma = 550$ km/s) containing 100 well-resolved disk galaxies 
inserted at $z=2$ (Dubinski 1998)
and a Virgo-scale cluster ($\sigma=800$ km/s) with 200 disks and 15
ellipticals inserted at $z=3$.  The main results of these
simulations are discussed below.

\section{Brightest Cluster Galaxies}

Most clusters have a giant elliptical or cD galaxy 
near their spatial and kinematic center. 
(They are generically known as brightest cluster galaxies or BCGs).
There have been various theories put forward for the origin 
of these galaxy giants
including
star formation in X-ray cooling flows (Fabian 1994), galactic cannibalism
and tidal destruction of small galaxies
(Richstone 1976; Ostriker \& Tremaine 1975), 
and early galaxy merging during the collapse of
cluster core in hierarchical structure formation (Merritt 1985).
There is little evidence of young stars in BCG's which refutes the cooling
flow idea and galactic cannibalism can only account for a small fraction of
the luminosity of a  BCG since merging is inefficient in virialized
clusters -- galaxies are moving fast and tidally truncated so dynamical
friction timescales are too long to allow much merging.

The idea that seems to work the best is rapid galaxy merging in a cosmological
hierarchy and has recently been illustrated by Dubinski (1998) in a
simulation of a poor cluster.
The 7 most massive galaxies in the collapsing cluster merge rapidly 
forming a BCG by $z\sim 1.0$ building up the bulk of its luminosity.
The BCG accretes a further 6 ``dwarf'' galaxies ($v_c \sim
100$ km/s) but they do not add much extra mass.  The structure and
kinematics of the resulting simulated BCG agree quantitatively with real
ones (cf. Fisher et al. 1995) 
with de Vaucouleurs light profiles ($r_e \approx 20$
kpc) and central velocity dispersions in the right range $\sigma \approx
350$ km/s).  Furthermore, the BCG displays the alignment effect: its
long axis is closely aligned with the long axis of the galaxy distribution
(Sastry 1968; Carter \& Metcalfe 1980).  This can be traced back to the
collapse of the cluster along the filament present in the cosmological 
initial conditions supporting conjectures that BCG's show alignment
correlations with large-scale structure (Binggeli 1982).

\section{Strong Tidal Interactions and Merging}

Moore et al. (1996) have pointed out that strong tidal interactions or
galaxy ``harassment'' in the
cluster environment play an important role in galaxy morphological
evolution.
In the simulations discussed here,
{\em most} galaxies show signs of tidal disturbance over their history.
Some major effects are the excitation of open spiral structures, warps and
even tidal tails in close encounters of individual galaxies with the 
growing central BCG.
Many galaxies are on orbits that take 
them within 100 kpc of the cluster center and the tidal fields within 
this radius are sufficient to obviously distort galaxies.
Isolated disturbed galaxies seen in
clusters may result in high speed encounters with
the parent cluster's central potential.
Higher resolution simulations of individual disks in orbit in fixed cluster
potentials clearly demonstrate this effect (Dubinski \& Hayes 1999).
Tidal heating of disks can be strong in this
environment and could account in part for the greater frequency of S0
galaxies in clusters as revealed by the morphology-density effect (Dressler
1980).

Galaxy-galaxy interactions also create disturbed galaxies
but this occurs mainly in bound pairs or groups which eventually merge forming
elliptical-like remnants.  It appears that the general origin of
elliptical galaxies in the cluster environment is the merging of sub-groups
in the hierarchy.  
Moore et al. (1998) claim that galaxy harassment may also
produce dwarf elliptical
galaxies but to be definitive higher resolution is needed since numerical
heating effects can be severe when simulating
galaxies over a Hubble time even when using 100K particles.
All of this work needs to be placed on a stronger, statistical foundation
with more simulations and better resolution.

\section{Intergalactic Tidal Debris}

Cluster tides are quite effective at stripping stars from galaxies 
and building up an intergalactic stellar population
and potentially account for the envelopes of the cD galaxies (Richstone
1976; Merritt 1985).  The recent discovery of intergalactic stars 
in the Virgo and Fornax clusters in the form of planetary nebulae
and red giant-branch stars (Theuns \& Warren 1996; Arnaboldi et al. 1996;
Feldmeier et al. 1998; Ferguson et al. 1998) along with the likelihood
of intergalactic globular clusters have revived interest
in the dynamics of the tidal-stripping process.

Analysis of the poor cluster simulation shows that
about 10\% of the stars are distributed diffusely throughout the cluster
with surface brightness dimmer than $\mu =26.5$ assuming an $M/L=5$
(Dubinski, Murali, \& Ouyed 1999).
The radial light distribution is approximately a continuation of a
deVaucouleurs profile out to $r=1$ Mpc from the center of the BCG.
The Virgo-cluster simulation shows similar results and there is the hint of
a cD envelope.

There are also a significant number of streams and swathes of
stars in the intracluster light that originate in stripping events of
spirals and ellipticals.
These streams often trace the orbits of galaxies
on radial orbits that have suffered a strong, tidal encounter with the
cluster center.  Examples of streams just forming in tidal encounters
have recently been detected in the Coma cluster (Gregg and West 1998).
Streams may be difficult to detect directly because of their low surface
brightness.
However, the contrast may increase when viewed in the $r-v_{los}$
phase-plane.
A kinematic survey of several hundred planetary nebulae and 
globular clusters in
the intergalactic space surrounding M87 in Virgo or NGC 1399 in Fornax 
may reveal coherent streams on top of a diffuse population.
New analysis shows how the kinematics of the tidal debris streams
present a new way to measure the 
gravitational potential of nearby clusters (Dubinski et al. 1999).

\section{Conclusions}

The simulations described here provide a detailed, quantitative way of
probing the dynamical evolution of galaxies in clusters.
Future work will concentrate on producing a simulated survey of clusters
covering a wide mass range in different cosmological models to improve
statistics on the BCGs and the elliptical population along with 
general effects of
tidal interactions on galaxy morphology over cosmic history.  In principle,
these effects depend strongly on the cosmological model so detailed
comparison with the observations will provide interesting constraints and
consistency checks on these models.

\acknowledgments

%resort alphabetically
%I am grateful to R. Carlberg, C. Murali, R. Ouyed,
%M. West, L. Sage, J. Schombert, S. Tremaine  and L. Hernquist 
%for comments and criticism on this work.
I acknowledge the Pittsburgh Supercomputing Center where some of these
calculations were done.

\end{document}